\documentclass[conference]{IEEEtran}
\IEEEoverridecommandlockouts
\usepackage{cite}
\usepackage{amsmath,amssymb,amsfonts}
\usepackage{algorithmic}
\usepackage{algorithm}
\usepackage{graphicx}
\usepackage{textcomp}
\usepackage{xcolor}
\usepackage{multirow}
\usepackage{siunitx}
\def\BibTeX{{\rm B\kern-.05em{\sc i\kern-.025em b}\kern-.08em
    T\kern-.1667em\lower.7ex\hbox{E}\kern-.125emX}}
\begin{document}

\title{A Nonlinear Hash{-}based Optimization Method for SpMV on GPUs}

\author{\IEEEauthorblockN
{
        Chen Yan$^{a,b}$, 
		Boyu Diao$^{a,b*}$
        \thanks{*The corresponding authors}, 
		Hangda Liu$^{a,b}$, 
		Zhulin An$^{a,b}$ 
		and Yongjun Xu$^{a,b}$
}

\IEEEauthorblockA{$^a$Institute of Computing Technology, Chinese Academy of Sciences, Beijing, China}

\IEEEauthorblockA{$^b$University of Chinese Academy of Sciences, Beijing, China}
\textit{\{yanchen23s, diaoboyu2012, liuhangda21s, anzhulin, xyj\}@ict.ac.cn}
}

\maketitle 

\begin{abstract}
\textbf{Sparse matrix-vector multiplication (SpMV) is a fundamental operation with a wide range of applications in scientific computing and artificial intelligence. However, the large scale and sparsity of sparse matrix often make it a performance bottleneck. In this paper, we highlight the effectiveness of hash-based techniques in optimizing sparse matrix reordering, introducing the Hash-based Partition (HBP) format, a lightweight SpMV approach. HBP retains the performance benefits of the 2D-partitioning method while leveraging the hash transformation's ability to group similar elements, thereby accelerating the pre-processing phase of sparse matrix reordering. Additionally, we achieve parallel load balancing across matrix blocks through a competitive method. Our experiments, conducted on both Nvidia Jetson AGX Orin and Nvidia RTX 4090, show that in the pre-processing step, our method offers an average speedup of 3.53 times compared to the sorting approach and 3.67 times compared to the dynamic programming method employed in Regu2D. Furthermore, in SpMV, our method achieves a maximum speedup of 3.32 times on Orin and 3.01 times on RTX4090 against the CSR format in sparse matrices from the University of Florida Sparse Matrix Collection.}

\end{abstract}


\section{Introduction}
Sparse matrix-vector multiplication (SpMV) has a wide range of applications, such as mathematical solutions for sparse linear equations \cite{b13}, iterative algorithm-solving processing \cite{b15}\cite{b25}, graph processing \cite{b9}\cite{b14}\cite{b24}, and weight calculations for forward and backward propagation in neural networks \cite{b3}\cite{b12}\cite{b17}\cite{b19}, etc. However, SpMV is actually the bottleneck for many algorithms. The sparse matrix used in SpMV has the following characteristics \cite{b4}: (1) Sparsity. On the one hand, sparse matrices contain a large number of zero elements. Calculating these zero elements is meaningless, so designing a storage format for sparse matrices is necessary. On the other hand, the distribution of nonzero elements in each sparse matrix is different, making it possible for optimization methods designed for one sparse matrix to achieve poor results on another sparse matrix. (2) Large Scale.  In practical applications, sparse matrices often represent distances between roads, weights between power nodes, contextual correlations, etc. These tasks all have the characteristic of large scale, the requirement of high bandwidth storage is much larger than what general devices actually have. Taking the GPU platform as an example, the large sparse matrix leads to a large number of global memory access, which seriously affects the SpMV speed \cite{b16}. 

The coordinate format (COO) records the value of each nonzero element and its row and column coordinates. This format is now widely used for storing sparse matrices. Early research on sparse matrix compression storage structures focused on how to save storage space more effectively. For example, the compressed sparse row format (CSR) is the most classic storage format for sparse matrices \cite{b18}\cite{b22}. In the CSR format, sparse matrices are stored row by row, and the $ptr$ array in CSR format records the position of nonzero elements at the beginning and end of each row. The col and data arrays are consistent with the COO format. It is evident that the CSR format effectively reduces storage space by compressing the number of recorded row coordinates. Algorithm 1 shows the parallel calculation of the SpMV. The Ellpack (ELL) format has advantages when the number of nonzero elements in each row is similar, and the Diagonal (DIA) format performs well in diagonal matrices. Each format achieves great performance in compression storage on a certain type of sparse matrix.

\begin{figure}[b]
    \label{csr SpMV}
    \renewcommand{\algorithmicrequire}{\textbf{Input:}}
    \renewcommand{\algorithmicensure}{\textbf{Output:}}
    \begin{algorithm}[H]
		\caption{CSR SpMV}
		\begin{algorithmic}[1]
		  \REQUIRE $csr\_ptr[],csr\_col[],csr\_data[],vect[],num\_of\_rows$ %%input
		  \ENSURE $output[]$    %%output
            \FOR{each $i \in [0,num\_of\_rows]$}
                \STATE {$j=csr\_ptr[i]$}
                \WHILE{$j < ptr[i+1]$}
                    \STATE {$sum += data[j] * vect[csr\_col[j]]$}
                    \STATE {$j ++$}
                \ENDWHILE
                \STATE {$output[i]=sum$}
            \ENDFOR
		\end{algorithmic}
    \end{algorithm}
\end{figure}

However, these methods focused more on how to reduce storage costs rather than optimize computational performance. In other words, the performance of matrix-matrix multiplication and matrix-vector multiplication using the above methods is inevitably constrained by sparse features\cite{b23}. How to balance the storage cost and the time overhead of sparse matrix-vector multiplication based on this format is currently the goal that researchers hope to achieve.

\begin{figure}[t]
\centering
\includegraphics[width=\linewidth]{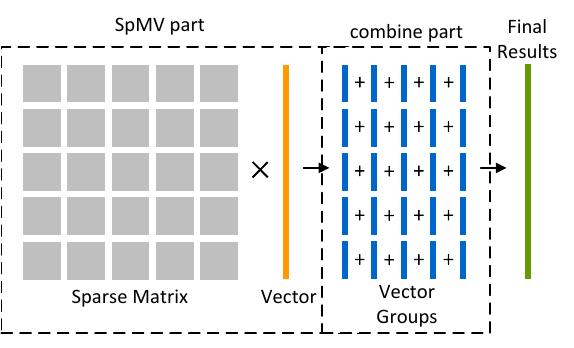}
\caption{Two-step SpMV. The matrix blocks and their corresponding vector segments are multiplied to obtain a set of vectors in the SpMV part. And the vectors in the same row are merged to get the final results in the combine part.}
\label{two-step SpMV}
\end{figure}

In practical applications, the scale of sparse matrices is often larger than that of the high-speed storage of parallel computing devices. This leads to frequent access to low-bandwidth storage during multiplication operations, further exacerbating the Memory Wall problem. Earlier proposed methods did not focus on this, so these methods have difficulty achieving good results in large-scale sparse matrix-vector operations. The 2D-partitioning of sparse matrices has been proven to be an effective method for solving random access to low-bandwidth memory \cite{b1}\cite{b10}\cite{b20}. Using the 2D-partitioning method, the SpMV calculation is divided into two parts as Fig. 1 shows. The first part multiplies matrix blocks by vector, called the "SpMV part", where each matrix block yields a vector segment. The second part involves combining the vectors that are located in the same row to obtain the final result vector, called the "combine part".

We mainly focus on the new problems that arise after the 2D-partitioning. After performing 2D-partitioning, the demand for balancing parallel loads transforms into balancing the computational loads between rows within matrix blocks and the computational loads between matrix blocks. 

Firstly, to balance the computational load between rows within matrix blocks, and improve SM utilization, reordering each row is used in many methods \cite{b21}. The sorting and dynamic programming methods \cite{b8}\cite{b7}achieve excellent results, but the cost of these methods cannot be ignored. Regardless of which method is used, it is necessary to first traverse the full matrix blocks to obtain the number of nonzero elements in each row, and then repeat multiple times based on this to obtain the final reordering result. Undoubtedly, the time cost is huge. In addition, although the method of load balancing through zero padding can achieve good results, the zero padding in format conversion results in uncertain matrix storage length, making it difficult to accelerate the format conversion steps in parallel. So, a lightweight method for SpMV optimization that balances preprocessing cost and computational speed is needed. We focus on the hash method, which maps the input to a new fixed interval according to specific rules. Therefore, we can use nonzero elements in each row as input and aggregate rows with similar numbers of nonzero elements through design rules. While ensuring the atomicity of the hashing process, each row within the matrix blocks can achieve parallel reordering. The hash method achieves significant speed improvement while ensuring high-quality reordering. We designed a new storage format named the hash-based partition format (HBP) based on nonlinear hashing to achieve parallelism at multiple granularities within and between matrix blocks.

Secondly, to balance the computational loads between matrix blocks, existing methods adopt the method of merging adjacent row or column matrix blocks, which balances loads by merging blocks with smaller computational loads into new blocks. However, this method not only requires additional storage to represent which blocks are merged but also destroys the advantage of 2D-partitioning, the limitation on the range of access vectors. We adopt the idea of "those who are capable work harder" in this regard, using the actual execution time of matrix blocks as the basis for scheduling. In the preprocessing step, the matrix is divided into two parts, one part is executed in parallel by threads, and the other part is competed for execution by threads that have completed the assigned tasks.

In summary, the contributions of this paper are:

1. A new hash-based partition storage format that enables parallel computation within a matrix block and between matrix blocks. We divide the matrix into the parallel execution part and the competing execution part, which balances the actual execution time efficiently.

2. A lightweight preprocessing step based on nonlinear hash. The hash transformation has a low time cost, effectively reducing the preprocessing time cost and avoiding thread divergence.

3. Our method achieves an average speedup of 3.53 times compared to the sorting approach and 3.67 times compared to the dynamic programming method employed in Regu2D. In SpMV, our method achieves a maximum speedup of 3.32 times on Nvidia Jetson AGX Orin and 3.01 times on RTX4090 against the CSR format in sparse matrices.

\section{Related Work}

\begin{figure*}[htbp]
\centering
\includegraphics[width=0.9\textwidth]{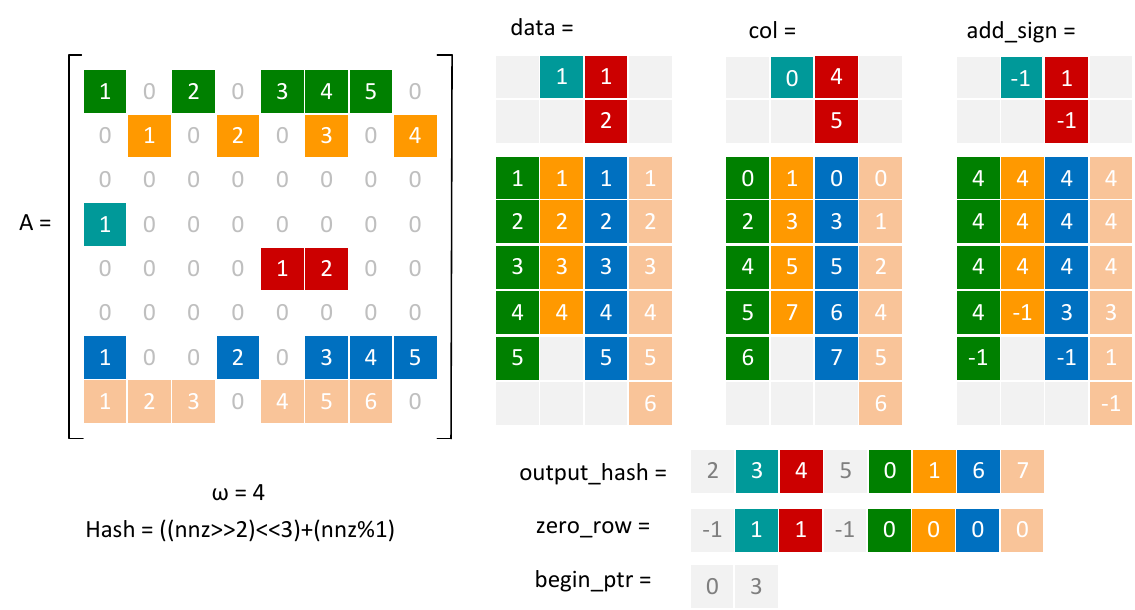}
\caption{An example of HBP format. We assume that the warp consists of 4 threads. $col$ and $data$ are stored in the order after hash transformation. $zero\_row$ and $begin\_ptr$ are used to find the first element that the current thread needs to calculate. $add\_sign$ represents the distance from the current element to the next element within the same row.}
\label{HBP}
\end{figure*}

In response to the problem of excessive time overhead in matrix-vector multiplication under traditional sparse matrix storage formats, relevant research teams have proposed many effective storage formats.

Liu et al. proposed the CSR5 \cite{b5} storage format, aiming to achieve ultimate load balance between parallel computing. The CSR5 storage format fills all nonzero elements in a sparse matrix into fixed-size matrix blocks one by one, with the length of the matrix block column direction equal to the size of the thread bundle. In this format, the number of computation operations performed by each thread is equal, thus achieving load balancing between threads.

Considering that CSR5 introduces a significant amount of storage cost to ensure parallel computing, the CVR \cite{b6} storage format was proposed, which aims to simulate the parallel computing process of threads. In the preprocessing step, the CVR format fills the nonzero elements in the sparse matrix in rows into a matrix with a column direction length equal to the size of the thread bundle. The difference is that in the CVR format, each row in the sparse matrix is only processed by one thread, so there is no need for merging operations after completing the sparse matrix calculation, and therefore no corresponding auxiliary array is required. 

These methods above only focus on parallel computing load balance, but do not take into account the frequent access to low-bandwidth storage overhead caused by the large-scale characteristics of sparse matrices.

Based on 2D-partitioning, Regu2D \cite{b7} employs dynamic programming within matrix blocks to balance the load. Additionally, for rows with similar numbers of nonzero elements, Regu2D pads these rows with zeros to ensure they are of exactly the same length. This is done to prevent divergence within each warp. This work also considers the balance issue between matrix blocks. For blocks with fewer nonzero elements, it merges them horizontally to achieve approximately equal computational loads among the matrix blocks. However, the introduction of zero-padding makes the storage length of the entire matrix variable. When preprocessing is performed in parallel, each thread needs to know the length of all previous storage and determine the write position accordingly. Therefore, dependencies exist among the matrix blocks, making it impossible to use parallel methods for acceleration.

\section{Method}

Previous works have demonstrated that designing a suitable sparse matrix storage format can effectively improve the computational speed of SpMV. Therefore, our work proposes a new storage format and corresponding SpMV calculation strategy, including the following parts.

\subsection{Hash-based Partition Format}

The Hash-based partition format (HBP) shown in Fig. 2 comprises six components: column indices of nonzero elements $col$, values of nonzero elements $data$, indices for accessing nonzero elements within a block $add\_sign$, indices of zero rows $zero\_row$, the storage positions of the first nonzero element in each block $begin\_nnz$, and a hash table $output\_hash$. To enhance locality in vector access, we first perform a 2D-partitioning of the sparse matrix. To ensure that each partitioned matrix block can be processed independently, we use $begin\_nnz$ to record the storage positions of the first nonzero element in each block. This array functionally equates to the $ptr$ array in CSR format, and during SpMV calculations, each block accesses the corresponding position in $begin\_nnz$ to obtain the starting point for computations within the current block. The HBP storage format stores the column indices of nonzero elements $col$ and the values of nonzero elements $data$ within the same block at adjacent locations. 

In each matrix block, we assign a warp of threads to each block to achieve parallelism within the matrix block. Each row needs to obtain its starting position. Since we have not introduced zero-padding, there are no corresponding nonzero element positions for zero rows. We use $zero\_row$ to record the positions of zero rows. If all elements in the row are zero, it is marked as -1; otherwise, it records the number of rows before the current row in the warp where the number of nonzero elements is zero. The zero-padding method is highly susceptible to the significant number of nonzero elements in some rows, leading to the storage of a large number of zero elements. In contrast, the storage cost of our method is fixed. $add\_sign$ is used to record the distance from one nonzero element to the next nonzero element in the same row within a matrix block. If the current element is the last nonzero element in the row, it is recorded as -1. The size of the array is equal to the number of nonzero elements. To avoid thread divergence within the matrix block, we use nonlinear hash mapping and reorder the rows. We employ $output\_hash$ to record the position of each row before the hash transformation, and the index of the hash table represents the actual execution order. The lengths of $zero\_row$ and $output\_hash$ are the same, equal to the number of rows in the sparse matrix multiplied by the number of column blocks.

The purpose of column partitioning the matrix is to ensure that during SpMV, access to the vector remains focused on localized segments, thereby leveraging spatial locality to accelerate memory access and address the issues posed by large-scale matrices. Therefore, the size of column partitioning is constrained by hardware resources. Although global planning would yield better results, the associated overhead cannot be ignored. We provide an example of selecting the size for 2D-partitioning. For Nvidia GPUs of compute capability over 2.x, by default the 48KB shared memory setting is used, we assume one warp of shared memory per block With a warp size of 32, implying each block contains 32 threads. In matrix-vector multiplication, the reusable part is a segment of the vector, stored using the double data type, where each element occupies 8 bytes. Therefore, each warp can store, on average, a vector segment with a maximum length of 6K in shared memory. Considering that shared memory will also need to store the identifiers of the matrix blocks to be executed, we allocate each warp to store a vector segment of length 4k. Thus, the size for 2D-partitioning in the column direction $M$ is set to 4096. Row partitioning of the matrix is intended to limit the scope of reordering. To balance the preprocessing speed and hash mapping effect, we set the partition size in the row direction $N$ to 512.

\subsection{Hash-based Reordering}

In previous methods that utilized reordering for optimization, dynamic programming was a widely recognized approach. However, this method required sorting based on the number of nonzero elements, introducing additional time overhead. We aim to achieve a similar effect in a more lightweight manner, and the hash function perfectly fits our needs.

In this paper, the hash function we employ can be abstracted into three parts: Aggregation, Dispersion, and Linear Mapping. The hash function shown in Fig. 3 takes the number of nonzero elements in each row within a block as input.

\begin{figure}
\centerline{\includegraphics[width=0.9\linewidth]{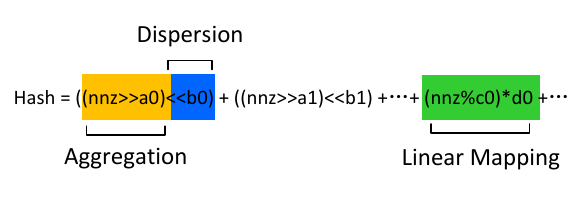}}
\caption{The format of nonlinear hash.}
\label{fig3}
\end{figure}

The aggregation maps rows with similar numbers of nonzero elements to the same position to achieve the purpose of aggregation. This part achieves a nonlinear mapping of the number of nonzero elements. The dispersion spreads the aggregated results to different positions in the hash table. Linear mapping makes fine adjustments within a smaller range to reduce collisions, thereby lowering the time cost of the entire hash process. In Fig. 3, $a$ and $c$ are dynamically determined based on the input matrix and sampled during program execution, while $b$ and $d$ are determined based on the size of the division in the row direction and are fixed before the program runs.

\begin{figure}
\centerline{\includegraphics[width=0.9\linewidth]{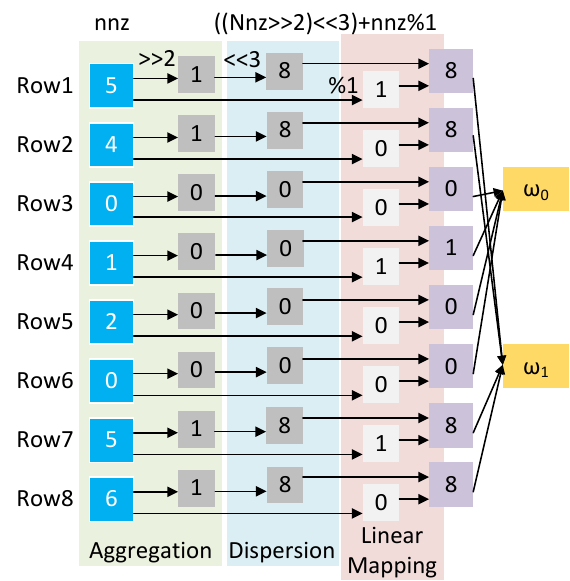}}
\caption{An example that shows the aggregation of Hash. Rows with fewer nonzero elements are aggregated after nonlinear hash mapping and computed by the warp of threads first.}
\label{fig4}
\end{figure}

In aggregation, rows with similar numbers of nonzero elements are mapped to the same value, achieving the purpose of aggregation. Taking the low-cost bit-shift operation as an example in Fig. 4, we assume $a$ is 2, this section can aggregate rows with the number of nonzero elements ranging from $4k$ to $4k+3$. Here, we artificially stipulate that the aggregation maps most numbers of nonzero elements to within the range of 0 to 8. Assuming there are four threads within each warp, where $\omega_0$ in Fig. 4 represents computations performed by the same warp at different time instances, it can be observed that during $\omega_0$, the hashing method aggregates rows with small computational loads. The computation time of a warp is determined by the thread with the longest computation time within it. Therefore, the smaller the difference in computational loads among threads within a warp, the more efficient the utilization of computational resources. During the sampling step, the program determines the values of $a$ and $c$ based on this range. To avoid the influence of extreme values on the results, we allowed the existence of a small number of rows that exceed 8 after mapping. These rows will be treated as rows assigned to 8 for subsequent calculations.

The dispersion disseminates the aggregate results to different positions in the hash table, with the mapping range not exceeding the column partition length of the current matrix block. The entire hash table is actually composed of smaller tables equal to the number of 2D-partitioning matrix blocks. Each row of each matrix block needs to correspond to a position in the hash table, so the total length of the hash table is equal to the number of matrix rows multiplied by the number of column blocks.

\begin{figure}[t]
    \label{data preparation}
    \renewcommand{\algorithmicrequire}{\textbf{Input:}}
    \renewcommand{\algorithmicensure}{\textbf{Output:}}
    \begin{algorithm}[H]
		\caption{Data Preparation}
		\begin{algorithmic}[1]
		  \REQUIRE $csr\_ptr[],csr\_col[],block\_col\_num,block\_row\_num$ %%input
		  \ENSURE $output\_hash[],nnz\_perrow[],begin\_nnz[]$    %%output
            \FOR{$thread\_id < num\_of\_rows$}
		  \STATE {$q = thread\_id \% M$}
            \STATE {$cur\_nnz=csr\_ptr[thread\_id]$}
            \STATE {$begin\_nnz[block\_m * M * block\_col\_num + q]=cur\_nnz$}
            \FOR{each $block\_n \in [0,block\_col\_num]$}
                \STATE {$basic=block\_m * M * block\_col\_num + block\_n * N + q$}
                \STATE {$j=cur\_nnz$}
                \WHILE{$csr\_col[j] < (block\_n + 1) * N  $}
                    \STATE {$nnz\_perrow[basic] ++$}
                    \STATE {$j++$}
                \ENDWHILE
                \STATE {$cur\_nnz=cur\_nnz + nnz\_perrow[basic]$}
                \STATE {$begin\_nnz[basic + N]=cur\_nnz$}
                \STATE {$hash$}
            \ENDFOR
            \ENDFOR
		\end{algorithmic}
    \end{algorithm}
\end{figure}

\begin{figure*}[ht]
\centering
\includegraphics[width=0.9\textwidth]{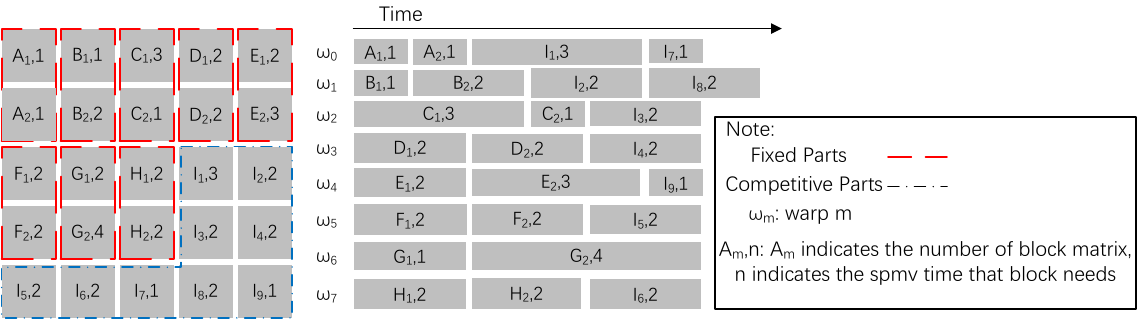}
\caption{An example of SpMV with mixed execution. Each matrix block is calculated by a warp. After a warp finishes computing its fixed parts, it will select the uncomputed matrix blocks from the competitive parts to perform SpMV.}
\label{fig5}
\end{figure*}

As matrix blocks become denser, the value of a will increase accordingly. In such cases, the aggregation would aggregate rows from a wider range together, which is obviously unreasonable. Therefore, we introduce the linear mapping component. The purpose of linear mapping is to fine-tune the mapping positions within a smaller range, which can include a term or a polynomial to achieve more precise mapping. The modulo operation exemplified in this paper can also be replaced by other methods such as bit-shifting.

Since the method of utilizing hash functions for nonlinear mapping mentioned above does not rely on the results of other rows within the matrix block, we can optimize using finer-grained parallelism. In the preprocessing step, we first compute the position of the first nonzero element for each row of the matrix block. Then, we sequentially search for the last nonzero element within the current block scope for that row. By doing so, we can count the number of nonzero elements in that row, apply hash mapping to obtain the reordered results and store the obtained results in a hash table. In this process, parallel computation is performed within a thread block on a row basis. To reduce the time cost of finding the position of the first nonzero element in each row, we assign each thread to process a complete row of the sparse matrix. In this way, the starting position of each block can be located using the ending position of the previous block. The implementation of this part is shown in Algorithm 2.

The aforementioned steps prepare the necessary data for format conversion. During the format conversion process, the granularity of parallel computation shifts from rows to matrix blocks. Within a matrix block, rows are grouped by thread warps. Hash operations ensure similar computational loads among rows within the group. Initially, rows within a thread bundle are traversed, and the number of zero rows is recorded in $zero\_row$. Subsequently, following column-major storage, we use $add\_sign$ to record the position from one element to the next within the same row, with its value equal to the current number of nonzero rows. After each nonzero element is added, we decrease the count of remaining nonzero elements in that row and check if all nonzero elements in the row have been added. If so, $add\_sign$ is updated to -1 to indicate that the current element is the last in the row, and the count of nonzero rows is decreased by one.

\subsection{Mixed Execution Allocation}

The aforementioned method does not consider the balance between matrix blocks. Previous work can be summarized as splitting blocks with dense elements or merging blocks with sparse elements \cite{b7}\cite{b8}. However, the scope of adjustment for such methods is limited due to the concentrated distribution of nonzero elements in sparse matrices in practical applications. For example, in matrix blocks with dense distributions of nonzero elements, the computational loads remain higher than those of blocks with very few or even zero nonzero elements even after splitting, which is quite common in sparse matrices. In this work, we no longer rely on the assumption that the number of nonzero elements is proportional to computation time, instead, we use actual execution time as the basis for scheduling.

In our work, the entire sparse matrix is divided into fixed parts and competitive parts. During SpMV, matrix blocks serve as the basic unit, with each block being computed by a warp of threads. In the fixed allocation parts, while ensuring an equal number of matrix blocks are assigned to each warp, we strive to allocate matrix blocks located on the same column to a single warp whenever possible. These matrix blocks access the same vector segments during SpMV computations, so we leverage shared memory to reduce the cost of memory access. The unassigned parts are classified as the competitive parts. Since the workloads of each matrix block in the fixed parts differ, leading to varying execution times, we allow warps that have completed their fixed allocations to atomically acquire matrix blocks from the competitive parts for computation. We employ ticket locks to regulate this process, enabling us to achieve load balancing across a broader scope through this method. The division of fixed and competitive parts bases on the scale of the input sparse matrix and the number of threads allocated. The entire division process requires only a minimal amount of time, and this cost is implicitly included in the pre-processing step.

Algorithm 3 demonstrates the process of parallel computation of SpMV on a matrix block. Here, $col$ represents the vector segment of length N corresponding to the current matrix block, which has been stored in shared memory. If the current row is a zero row, zeros are directly written to the corresponding positions in the output. Otherwise, the computation loops until the $add\_sign$ value is less than zero, indicating the end of the current row. The positions where values are written are those before the hash transformation.

\begin{figure}[!t]
    \label{Hash-based SpMV in a single matrix block}
    \renewcommand{\algorithmicrequire}{\textbf{Input:}}
    \renewcommand{\algorithmicensure}{\textbf{Output:}}
    \begin{algorithm}[H]
		\caption{Hash-based SpMV in a Single Matrix Block}
		\begin{algorithmic}[1]
		  \REQUIRE $col[],data[],add\_sign[],output\_hash[],zero\_row[],$\\
          $begin\_ptr[],vect[]$ %%input
		  \ENSURE $output[]$    %%output
		  \STATE {$q = thread\_id \% M$}
            \STATE {$a = block\_m * block\_col\_num * M + block\_n * N$}
            \FOR{$i < M$}            
            \IF{$zero\_row[a + i] == -1$}
            \STATE {$output[a + output\_hash[begin + i]] = 0$}
            \ELSE
            \STATE {$j=begin\_ptr[(block\_m*block\_col\_num+block\_n)*(M/\omega)] + q - zero\_row[a + i]$}
            \STATE {$sum = 0$}
            \WHILE{$add\_sign[j] > 0$}
                \STATE {$sum += data[j] * vect[col[j]\% N]$}
                \STATE {$j+=add\_sign[j]$}
            \ENDWHILE
            \STATE {$output[a + output\_hash[begin + i]] = sum $}
            \ENDIF
            \ENDFOR
		\end{algorithmic}
    \end{algorithm}
\end{figure}

\begin{figure*}[htbp]
    \begin{minipage}{0.32\textwidth}
		\centerline{\includegraphics[width=\textwidth]{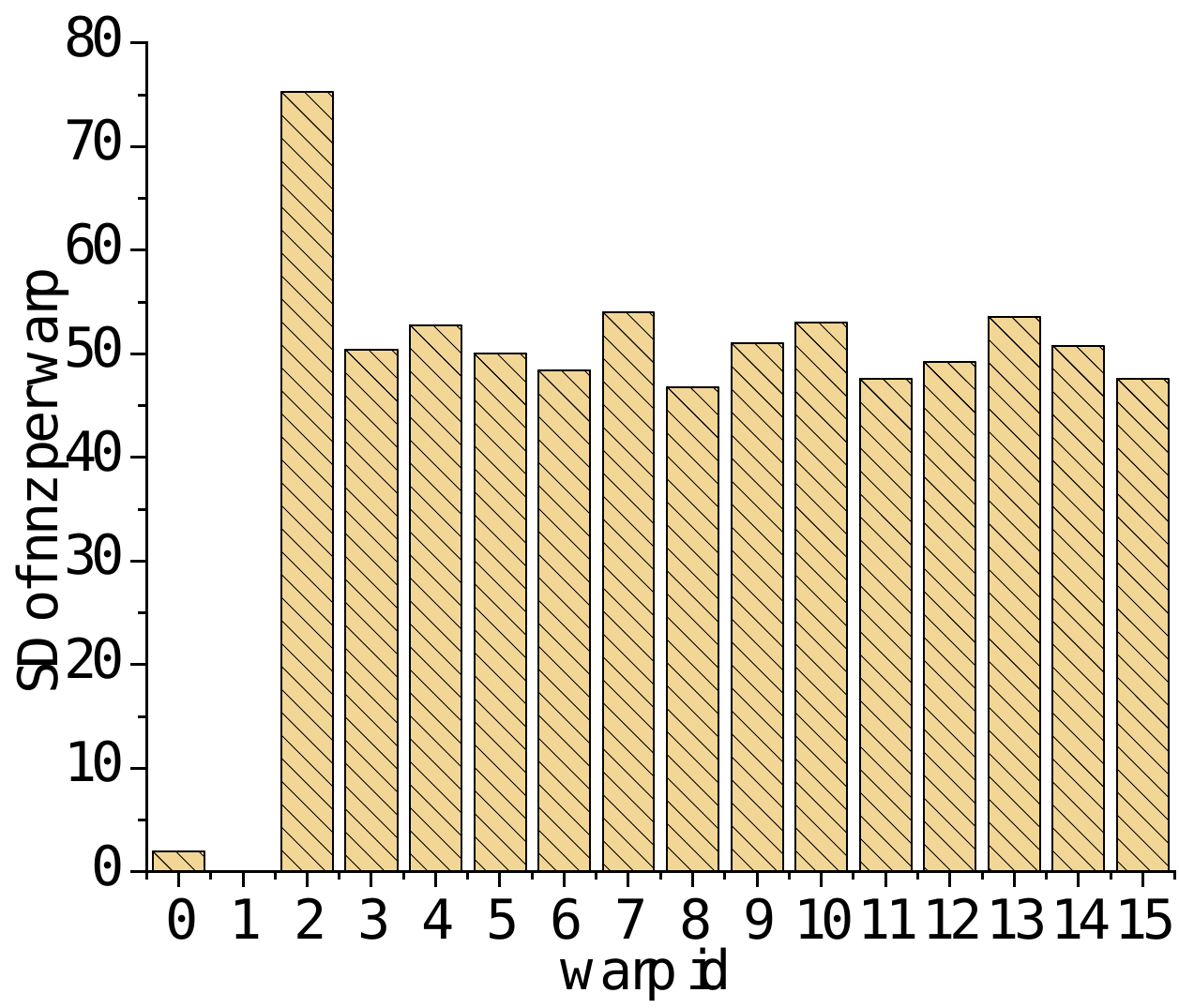}}
		\centerline{\includegraphics[width=\textwidth]{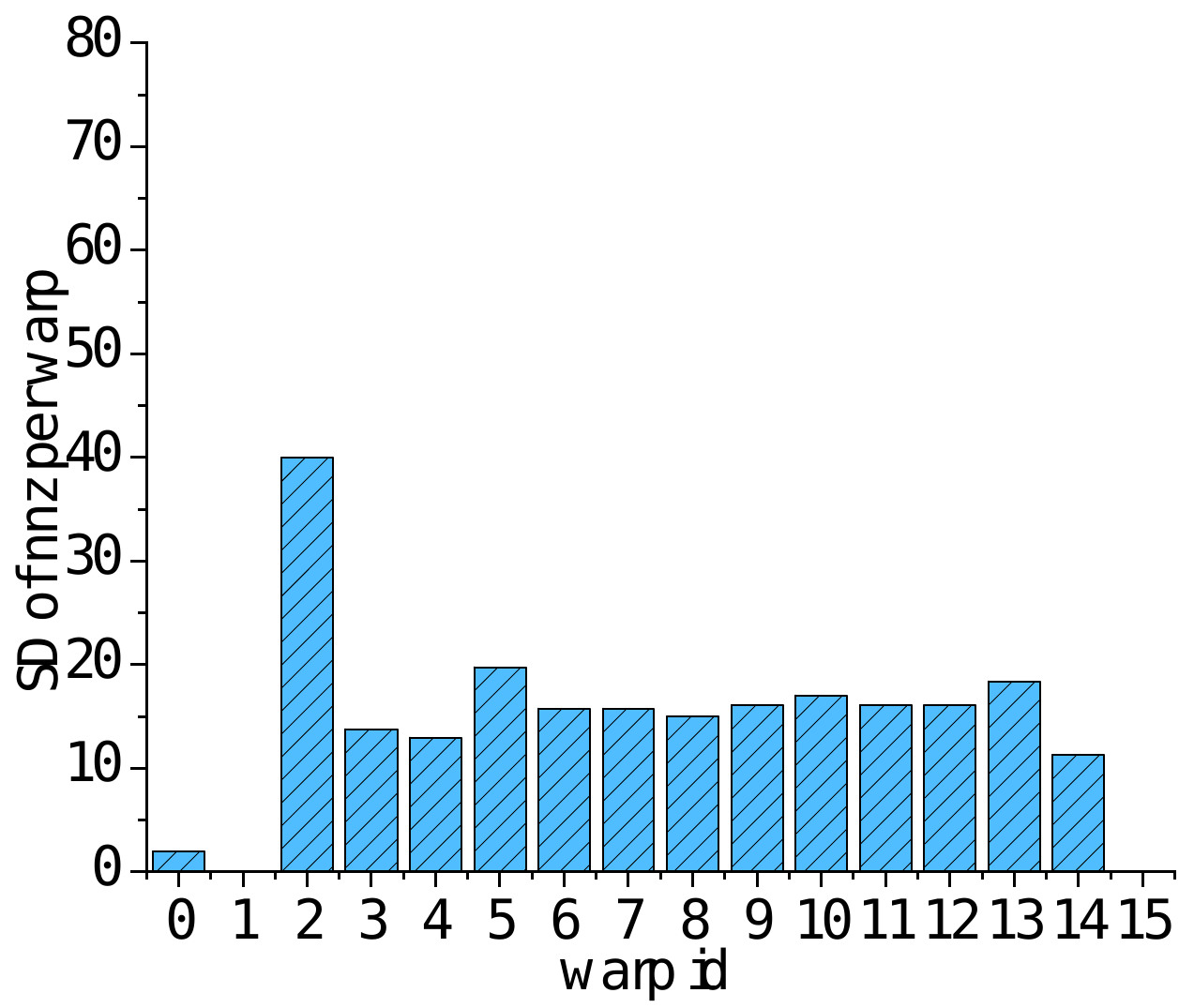}}
        \centerline{ASIC\_680k}
	\end{minipage}
    \begin{minipage}{0.32\textwidth}
		\centerline{\includegraphics[width=\textwidth]{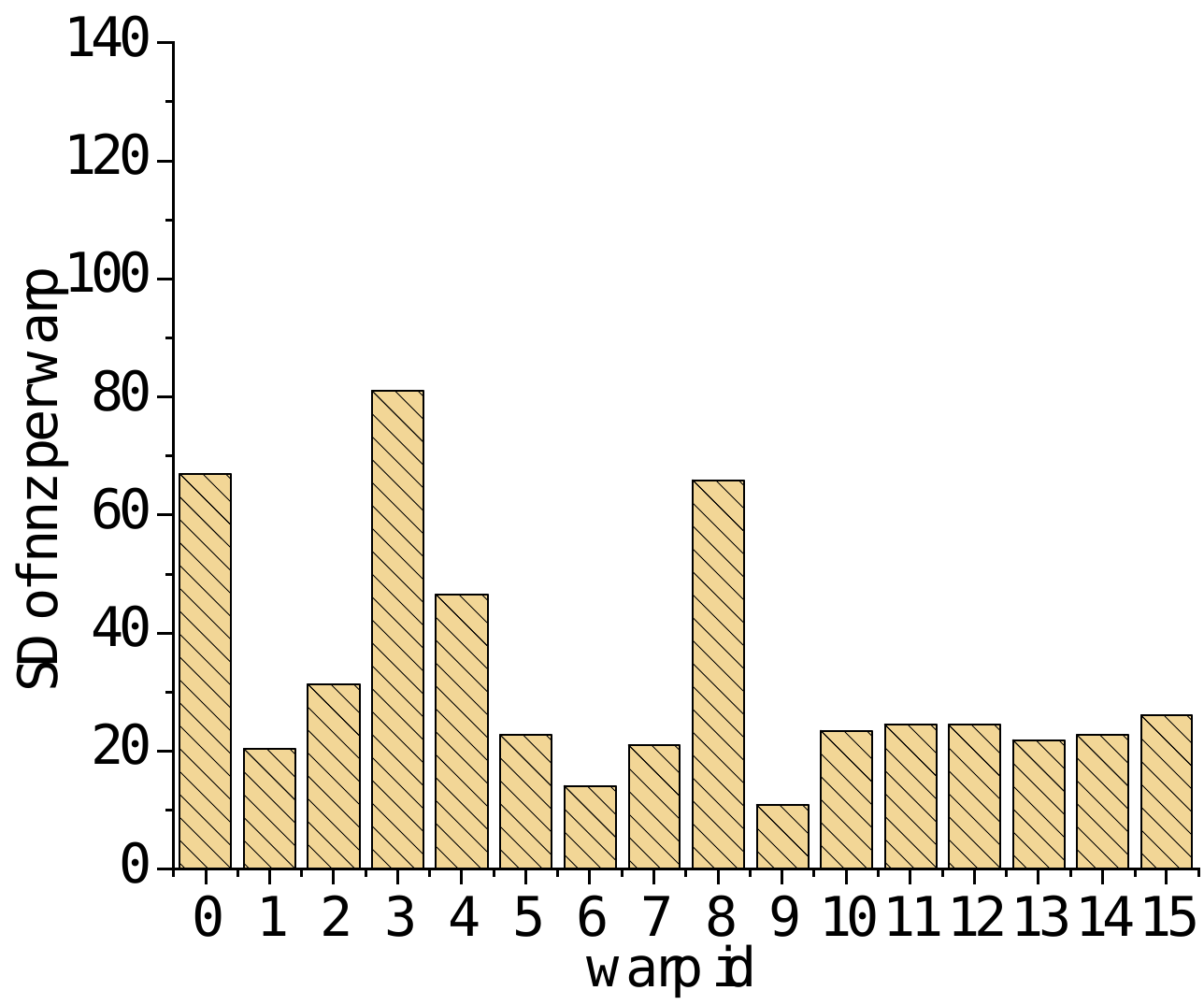}}
		\centerline{\includegraphics[width=\textwidth]{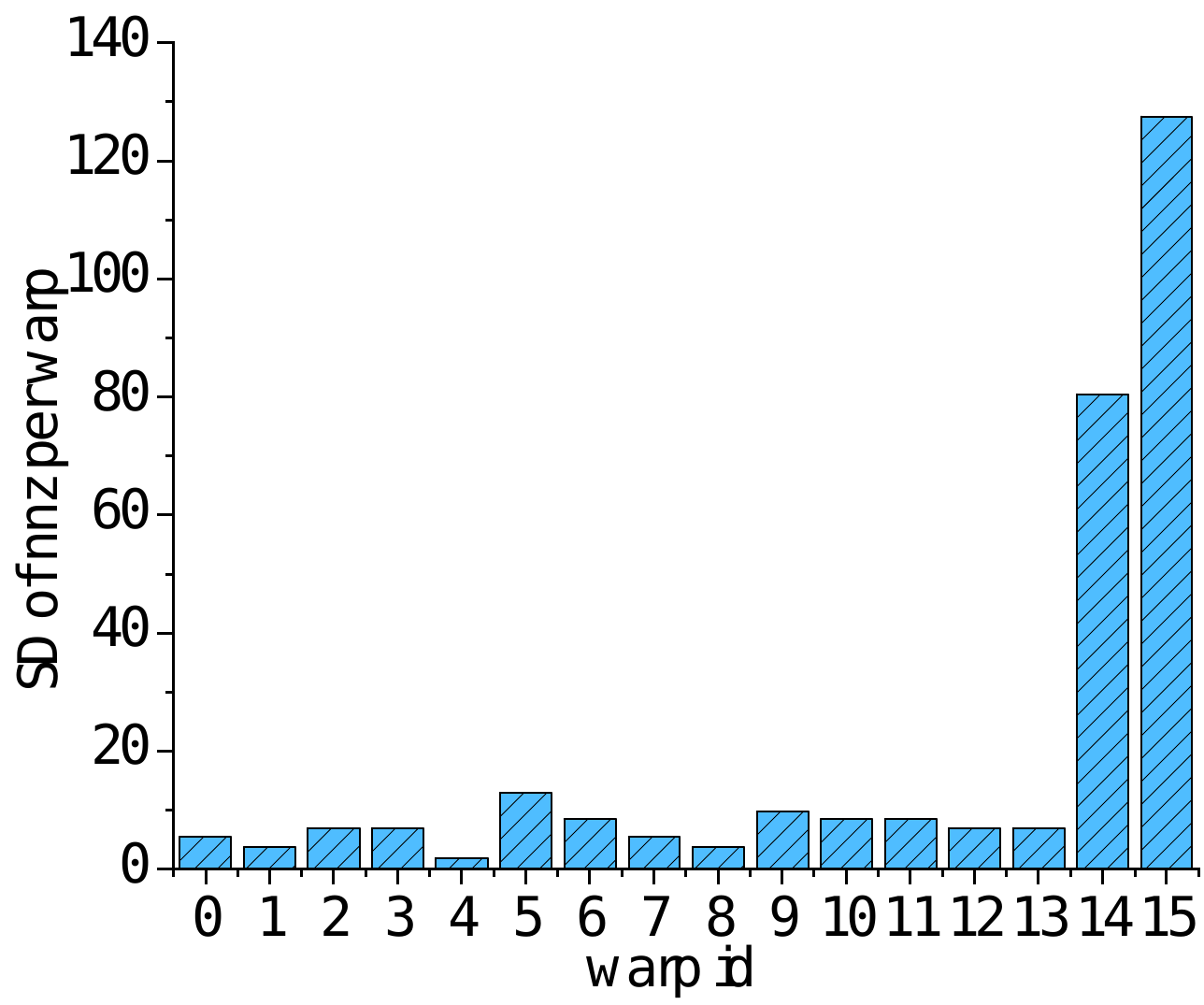}}
        \centerline{kron\_g500-logn18}
	\end{minipage}
    \begin{minipage}{0.32\textwidth}
		\centerline{\includegraphics[width=\textwidth]{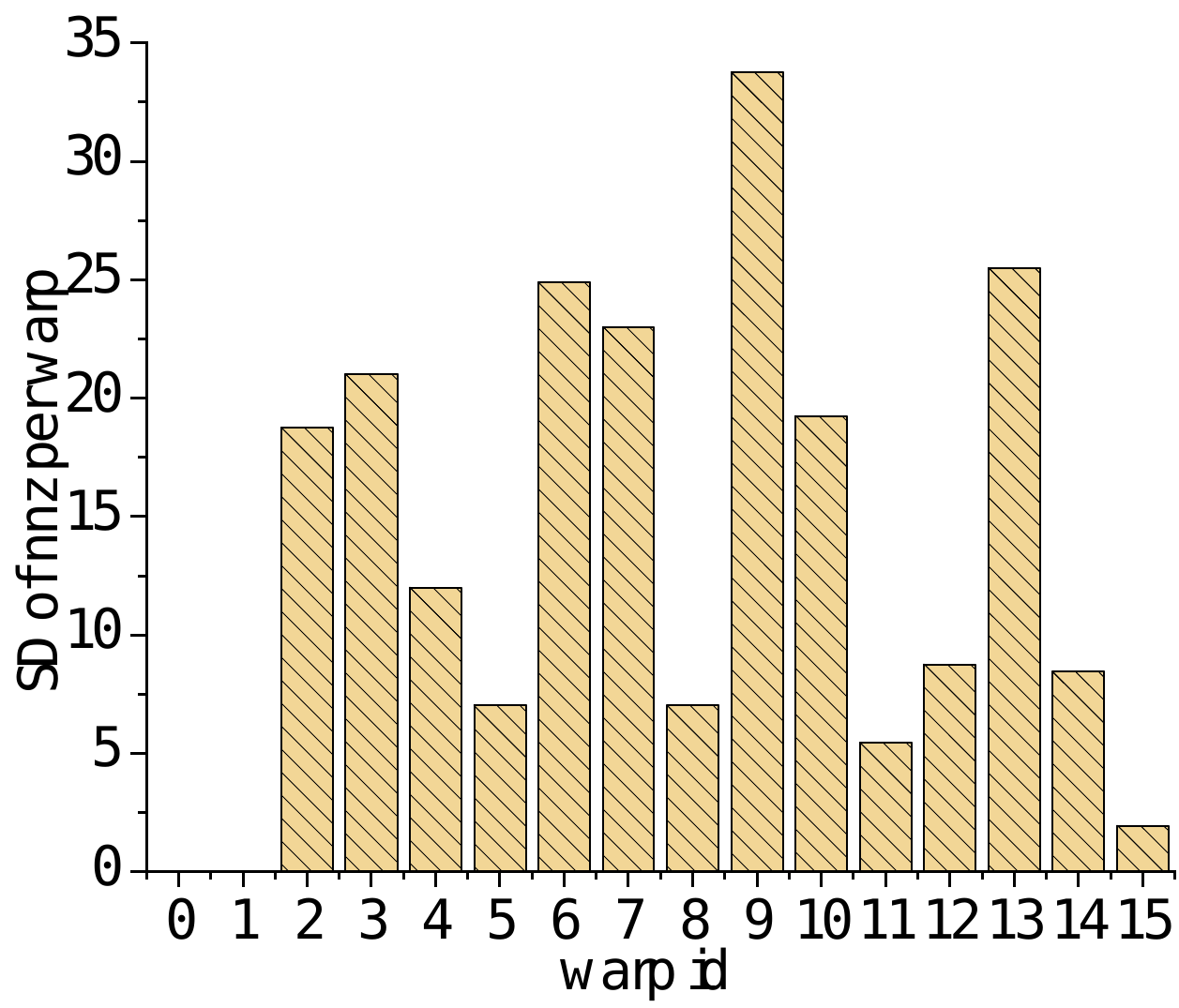}}
		\centerline{\includegraphics[width=\textwidth]{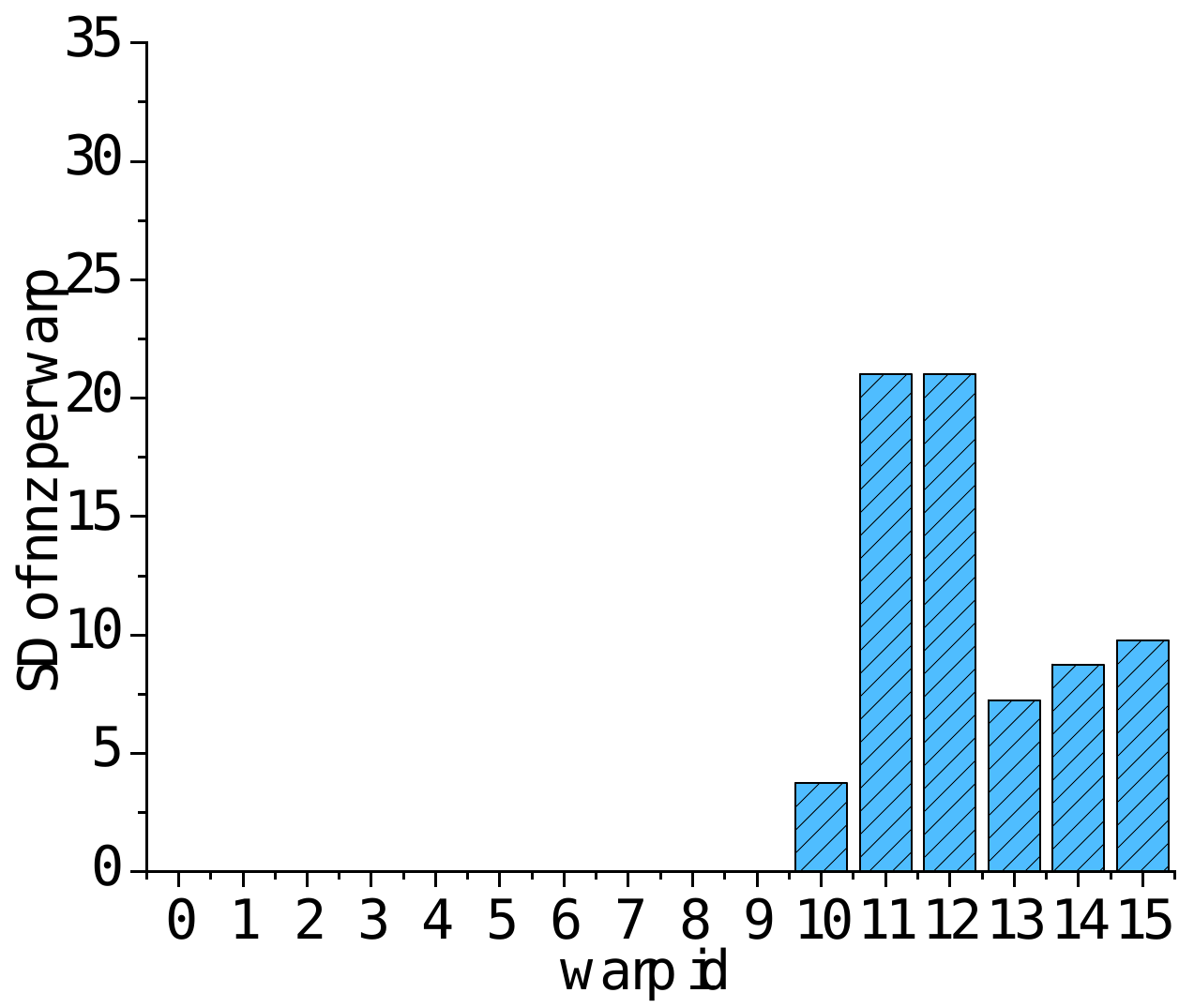}}
        \centerline{nxp1}
	\end{minipage}
    \\
    \centering
    \begin{minipage}{0.33\textwidth}
		\centerline{\includegraphics[width=\textwidth]{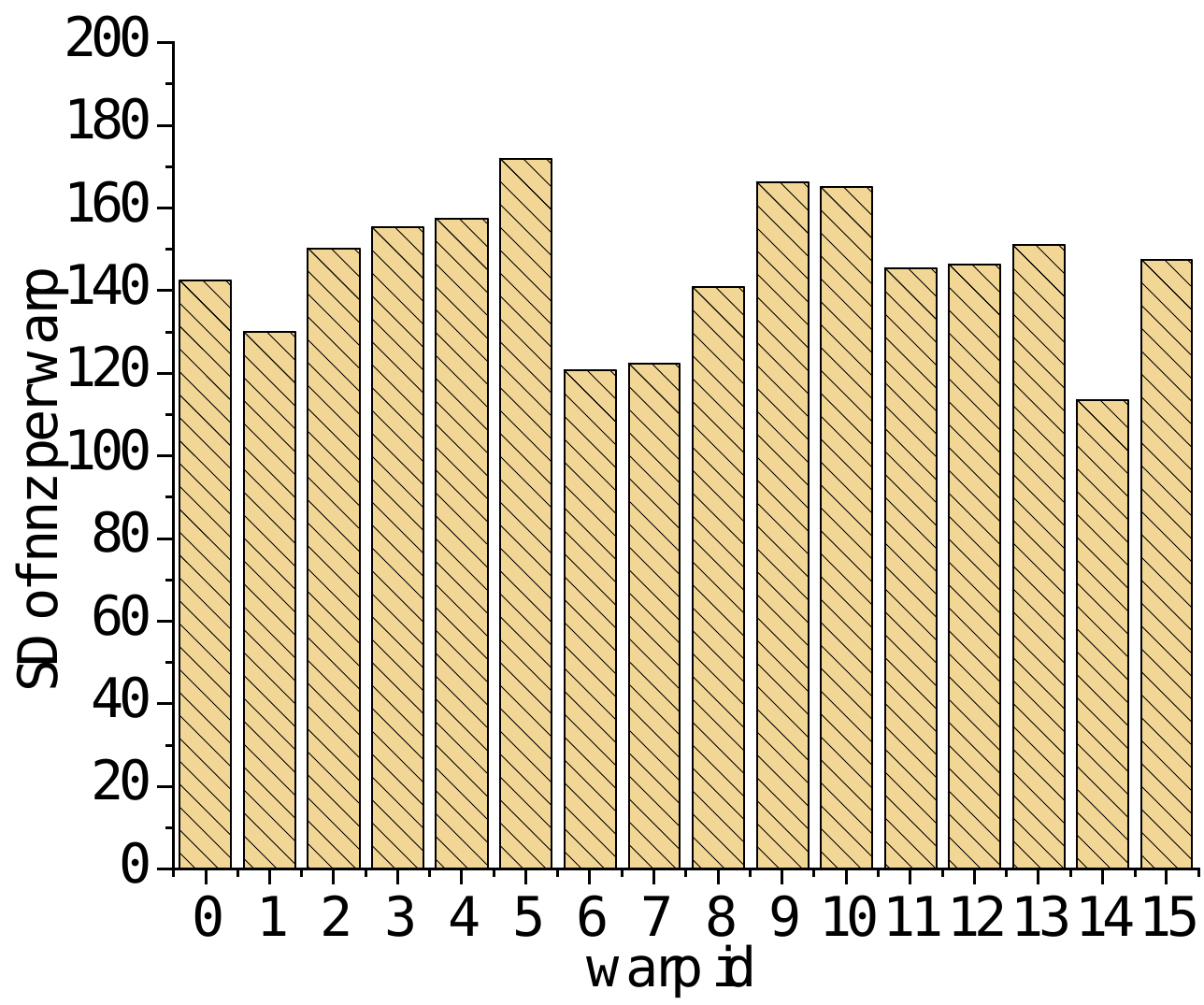}}
		\centerline{\includegraphics[width=\textwidth]{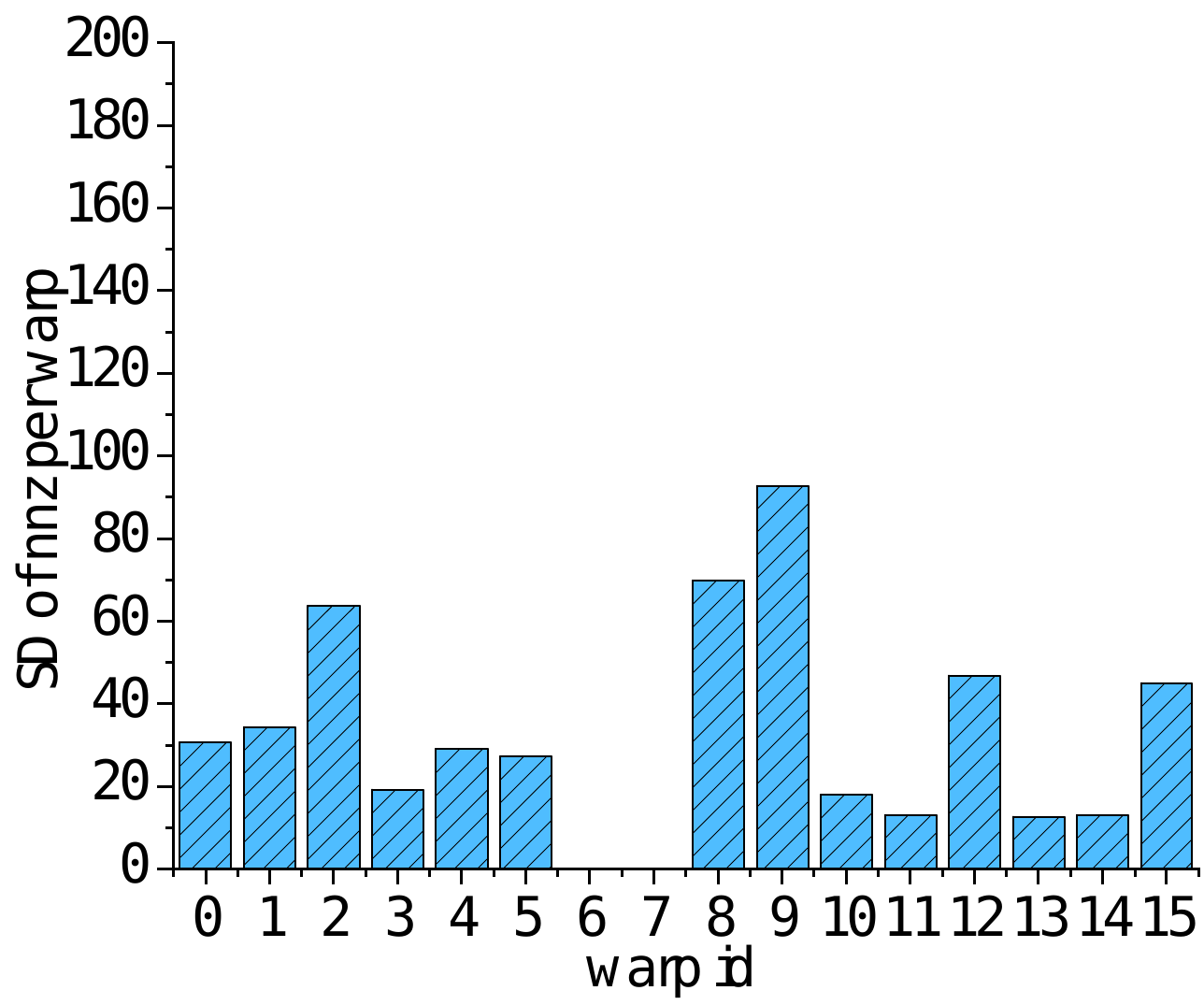}}
        \centerline{ohne2}
	\end{minipage}
    \begin{minipage}{0.33\textwidth}
		\centerline{\includegraphics[width=\textwidth]{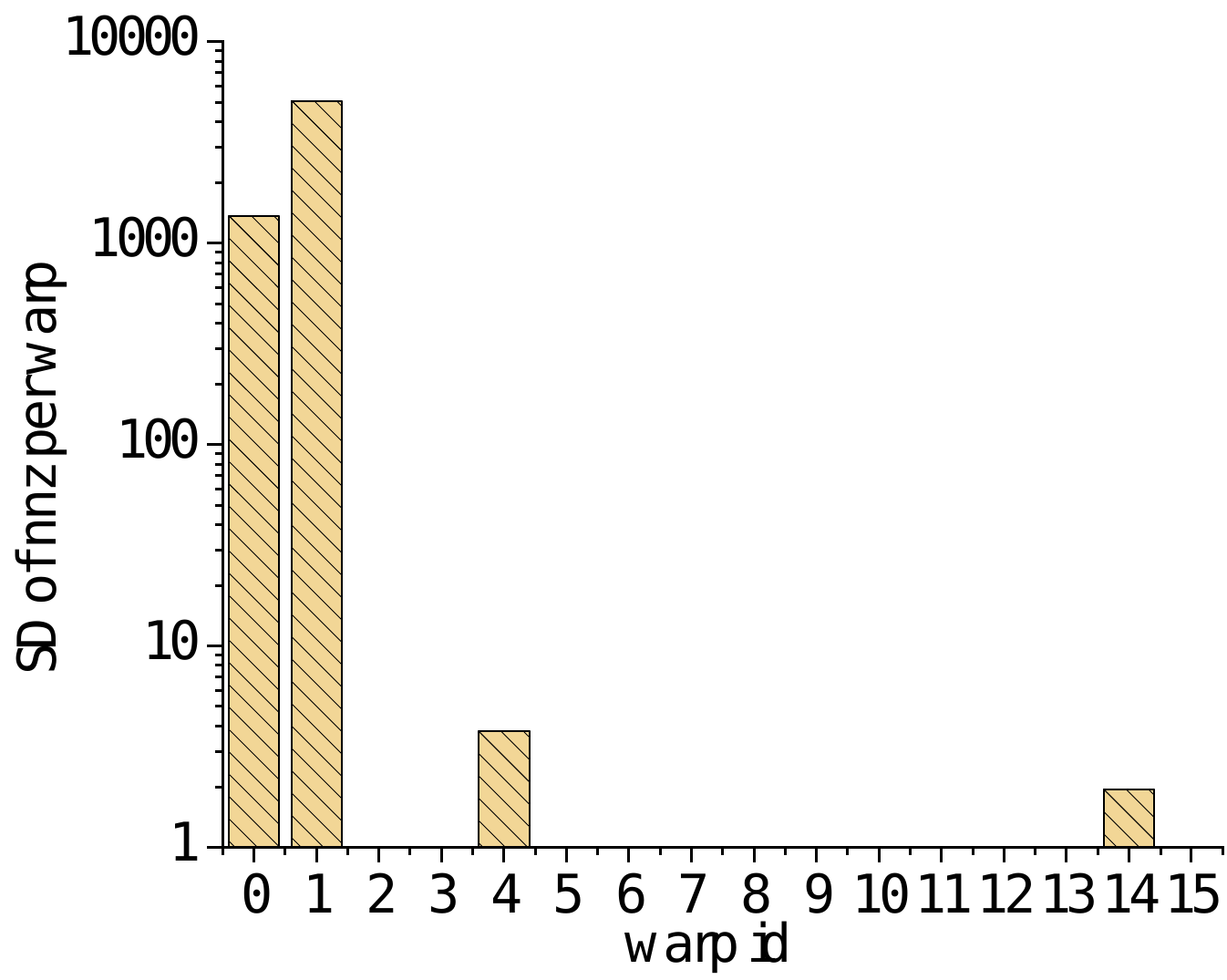}}
		\centerline{\includegraphics[width=\textwidth]{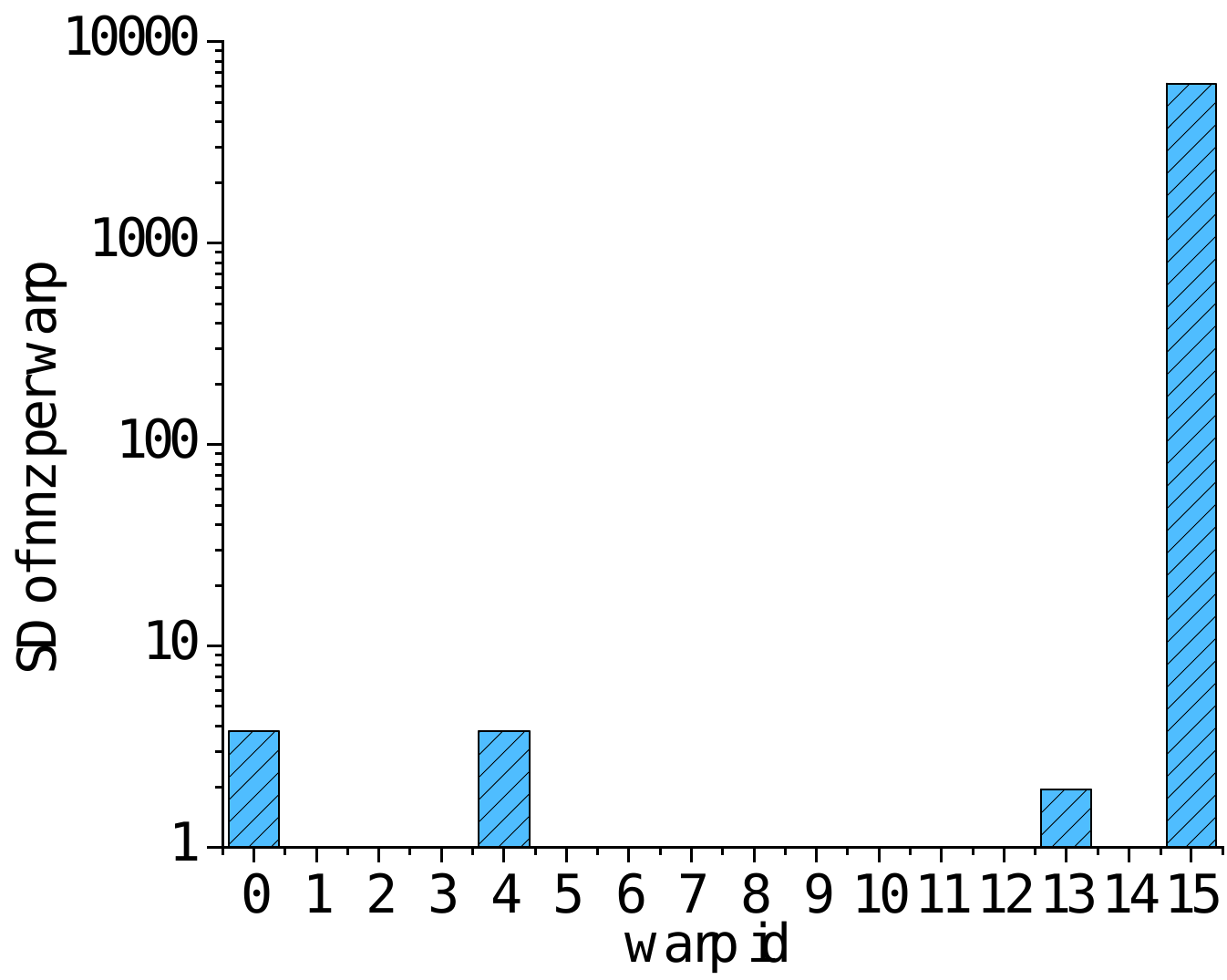}}
        \centerline{rajat30}
	\end{minipage}
    \caption{The ordinate represents the standard deviation of nonzero elements in each row within each warp in a matrix block. The yellow represents the original standard deviation of each group within the matrix block, blue represents the standard deviation after the hashing transformation. In our partitioning, each matrix block contains 16 groups, and the number of rows in each groups is equal to the size of a warp. The rows within the same group are executed simultaneously by the threads within the warp.}
    \label{fig6}
\end{figure*}

\section{Evaluation}
We conduct experiments on both NVIDIA Jetson AGX Orin 64GB and RTX 4090. The Jetson AGX Orin is a high-performance AI supercomputer released by NVIDIA, equipped with a 2048-core NVIDIA Ampere architecture GPU featuring 64 Tensor Cores. The RTX 4090 has up to 16384 cores, excellent computing performance, and is equipped with 24GB of GDDR6X high-speed video memory, providing high bandwidth and high-speed data transmission. These two devices represent large graphic memory and high computing power, respectively. To verify the effectiveness of our optimizations, we use over 10 sparse matrices from the University of Florida Sparse Matrix Collection \cite{b2}, which covers most sparse matrices used in previous works. Table 1 lists the info of the sparse matrices we used.

\begin{table}[b]
\caption{Test Sparse Matrices}
\begin{center}
\begin{tabular}{|c|c|c|c|c|}
\hline
\textbf{Matrix Id} & \textbf{Name}& \textbf{Dimensions}& \textbf{Nnz} \\
\hline
m1 & ASIC\_320k & 321K $\times$ 321K & 1.9M \\
\hline
m2 & ASIC\_680k & 682K $\times$ 682K & 3.8M \\
\hline
m3 & barrier2-3 & 113K $\times$ 113K & 2.1M \\
\hline
m4$^{\mathrm{*}}$ & kron\_g500-logn18 & 262K $\times$ 262K & 21.1M \\
\hline
m5$^{\mathrm{*}}$ & kron\_g500-logn19 & 524K $\times$ 524K & 43.5M \\
\hline
m6$^{\mathrm{*}}$ & kron\_g500-logn20 & 1.0M $\times$ 1.0M & 89.2M \\
\hline
m7$^{\mathrm{*}}$ & kron\_g500-logn21 & 2.0M $\times$ 2.0M & 182.0M \\
\hline
m8$^{\mathrm{*}}$ & mip1 & 66K $\times$ 66K & 10.3M \\
\hline
m9 & nxp1 & 414K $\times$ 414K & 2.7M \\
\hline
m10 & ohne2 & 181K $\times$ 181K & 6.9M \\
\hline
m11 & rajat21 & 411K $\times$ 411K & 1.8M \\
\hline
m12 & rajat24 & 358K $\times$ 358K & 1.9M \\
\hline
m13 & rajat29 & 643K $\times$ 643K & 3.8M \\
\hline
m14 & rajat30 & 643K $\times$ 643K & 6.2M \\
\hline
\multicolumn{4}{l}{$^{\mathrm{*}}$Symmetric matrix.}
\end{tabular}
\label{tab1}
\end{center}
\end{table}

\subsection{Parallel Load Balancing within Matrix Blocks}\label{AA}

The parallel load balancing within matrix blocks is achieved through nonlinear hashing. To evaluate the effectiveness of nonlinear hashing, we use the standard deviation of nonzero elements per warp of rows within a matrix block as a metric. A large standard deviation indicates great variation in the number of nonzero elements among rows within the same warp, implying that more computational resources are wasted. We selected matrix blocks with rows not entirely consisting of zeros from various sparse matrices for comparison. The standard deviation of the number of nonzero elements per warp across initial matrix blocks after 2D-partitioning and those after undergoing nonlinear hashing transformation is illustrated in Fig. 6. Here, the abscissa represents the workload computed by a warp at the same time. We used a row partition size of 512 and a warp size $\omega$ of 32, so that each matrix block is divided into 16 groups. The ordinate represents the standard deviation of the number of nonzero elements per warp of rows within the matrix block. In Fig. 6, yellow represents the original standard deviation of each group within the matrix block, while blue represents the standard deviation after the hashing transformation. Compared to the original sparse matrix blocks, using nonlinear hashing reduces the standard deviation of the matrix blocks by 42\% ($kron\_g500-logn18$), 79\% ($ASIC\_680k$), 67\% ($nxp1$), 78\% ($ohne2$), and 5\% ($rajat30$), respectively.

In Fig.6, the distribution of the number of nonzero elements within each group in the original sparse matrix varies significantly. For example, the standard deviation of each group in the $ASIC\_680k$ and $ohne2$ matrix blocks is obviously high. After hash mapping, the standard deviation is significantly reduced, which means we balanced the computational load among threads within the warp, thereby improving thread utilization. Due to the excessive number of nonzero elements in some rows, and these rows are not sufficient to fill a warp, low thread utilization is unavoidable. However, hash mapping still successfully maps rows with large computational loads originally located under different warps to the same warp, in order to achieve an overall increase in computation speed.

\subsection{Evaluation of the Preprocessing Step}

\begin{figure}[b]
\centerline{\includegraphics[width=0.9\linewidth]{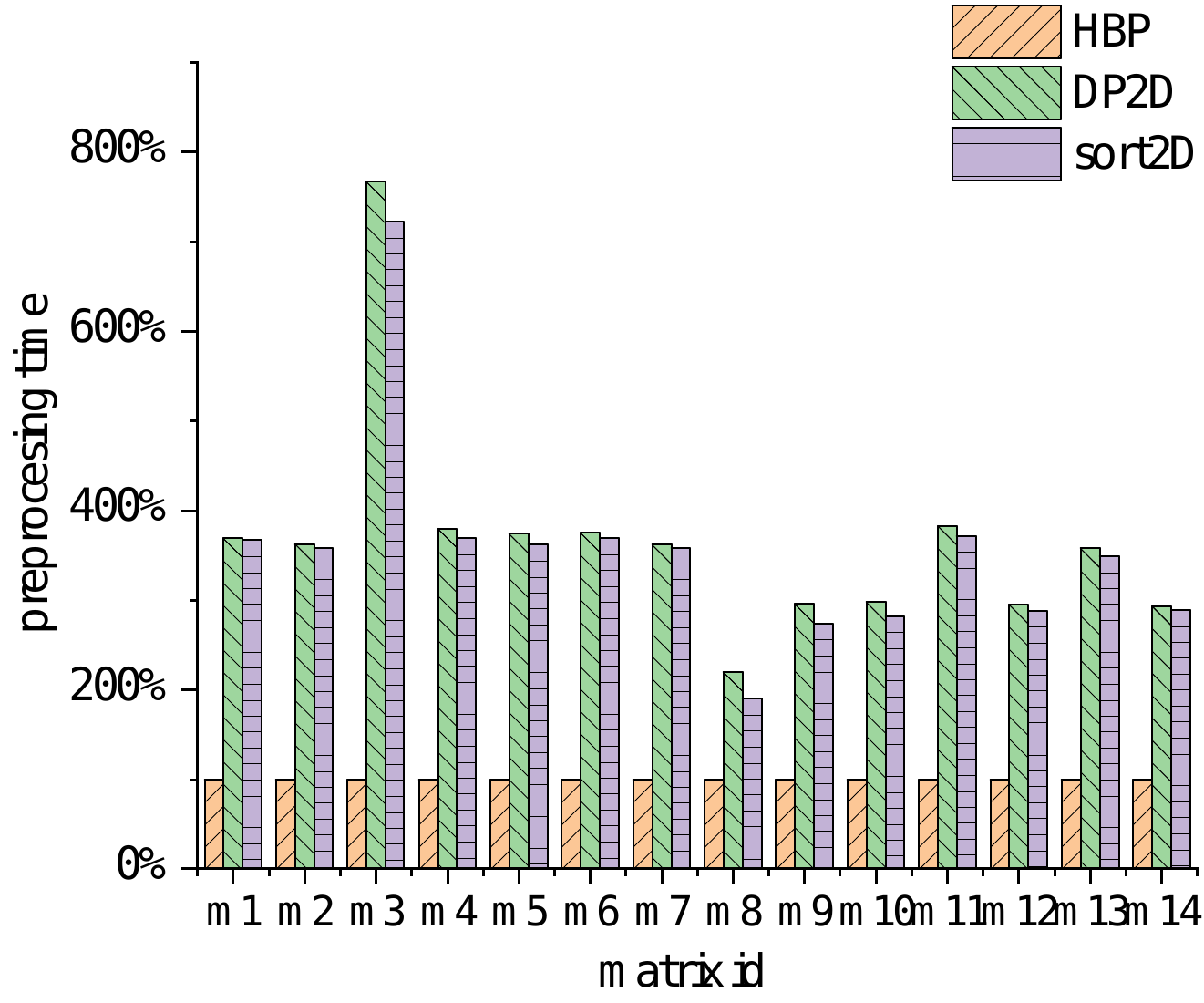}}
\caption{Evaluation of preprocessing on Nvidia Jetson AGX Orin. The ordinate represents the proportion of time that other methods required for the preprocessing step compare to the HBP method we proposed.}
\label{fig7}
\end{figure}

To evaluate preprocessing costs, we choose the basic sorting method (sort2D) and the dynamic programming approach used in the Regu2D preprocessing step (DP2D). Since the preprocessing method of Regu2D involves zeros padding into the storage matrix to enhance memory access efficiency, this approach necessitates a serial execution of the preprocessing step. Such a comparison would be meaningless; therefore, we only use its dynamic programming method as a benchmark.

As shown in Fig. 7, our method achieves a maximum speedup of 7.67x (with an average speedup of 3.67x) compared to the DP2D method and a maximum speedup of 7.23x (with an average speedup of 3.53x) compared to the sort2D method. This attribute to the nonlinear hashing method we employed. The DP2D method incorporates a sorting step, which necessitates obtaining the number of nonzero elements in each row of the entire matrix block. However, the sorting process is not conducive to parallel acceleration, making sorting a bottleneck in the preprocessing step. The nonlinear hash mapping we used addresses this issue. 

\begin{figure}[htbp]
\centerline{\includegraphics[width=0.9\linewidth]{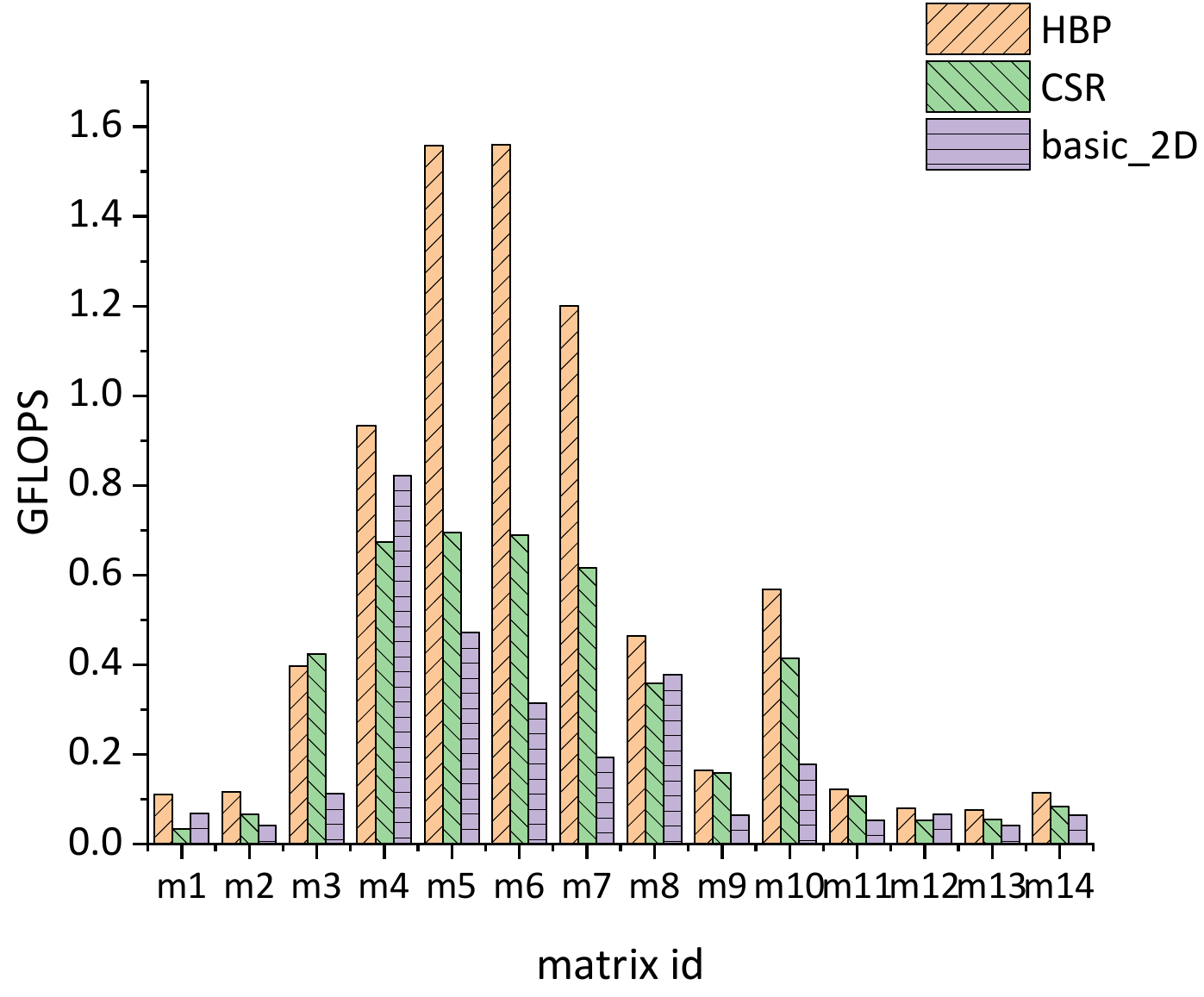}}
\caption{Evaluation of SpMV on Nvidia Jetson AGX Orin. The number of necessary computations to perform SpMV is constant in a certain matrix. We obtain GFLOPS by dividing this number of computations by the sum of SpMV time and combine time. A higher GFLOPS indicates faster actual execution time and more efficient utilization of resources.}
\label{fig8}
\end{figure}

\subsection{Evaluation of the SpMV Step}

We compared the SpMV speed of our method with that of CSR format and 2D-partitioning on Nvidia Jetson AGX Orin and Nvidia RTX 4090. The results are shown in Fig. 8 and Fig. 10. The number in the y-coordinate is the GFLOPS calculated by $G = 2 \cdot nnz/t$ where $nnz$ is the number of nonzero elements and $t$ is the execution time. The higher GFLOPS represents a more efficient utilization of GPUs.

As shown in Fig. 8, our method achieves a maximum speedup of 3.32x (with an average speedup of 1.64x) compared to the CSR method and a maximum speedup of 6.17x (with an average speedup of 2.68x) compared to the 2D-partitioning method on Nvidia Jetson AGX Orin. From Fig. 8, we can see that:

\begin{itemize}
\item On the $m4$ and $m8$ matrices, the SpMV computation speed is affected by the issue of scattered vector access locations. Therefore, our method and the 2D-partitioning method outperform the CSR method. Furthermore, our method achieves a good balance between the workload among warps and within threads within warps, leading to better performance compared to the 2D-partition method.
\item Using the 2D-partitioning method only cannot achieve better optimization effects on most matrices. As the size of sparse matrices increases, the computational cost incurred by 2D-partitioning, especially the time consumed by the merge step in SpMV increases accordingly. We have collected statistics on the time overhead for performing SpMV computations and the merge step as the matrix size grows. The results are shown in Fig. 9. The optimization of our method for the SpMV step is evident. However, when the matrix size further increases, the merge step in the 2D-partitioning method becomes the primary factor limiting computation speed. 
\end{itemize}

\begin{figure}[t]
\centerline{\includegraphics[width=0.9\linewidth]{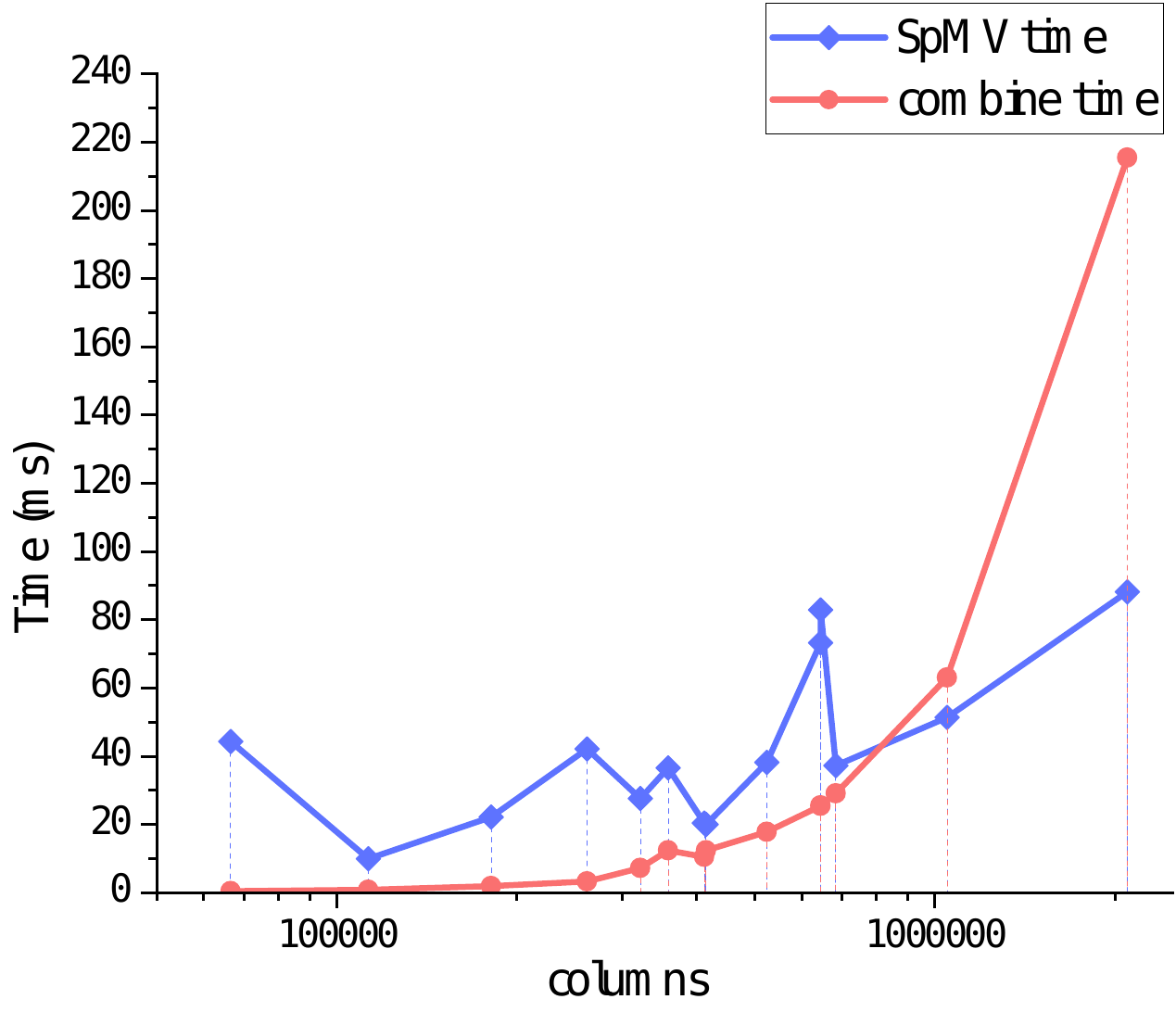}}
\caption{The time of SpMV and combine parts on Nvidia Jetson AGX Orin. As the size of the matrix increases, the growth rate of the time required for the combine part significantly exceeds that of the SpMV part.}
\label{fig9}
\end{figure}

\begin{figure}[htbp]
\centerline{\includegraphics[width=0.9\linewidth]{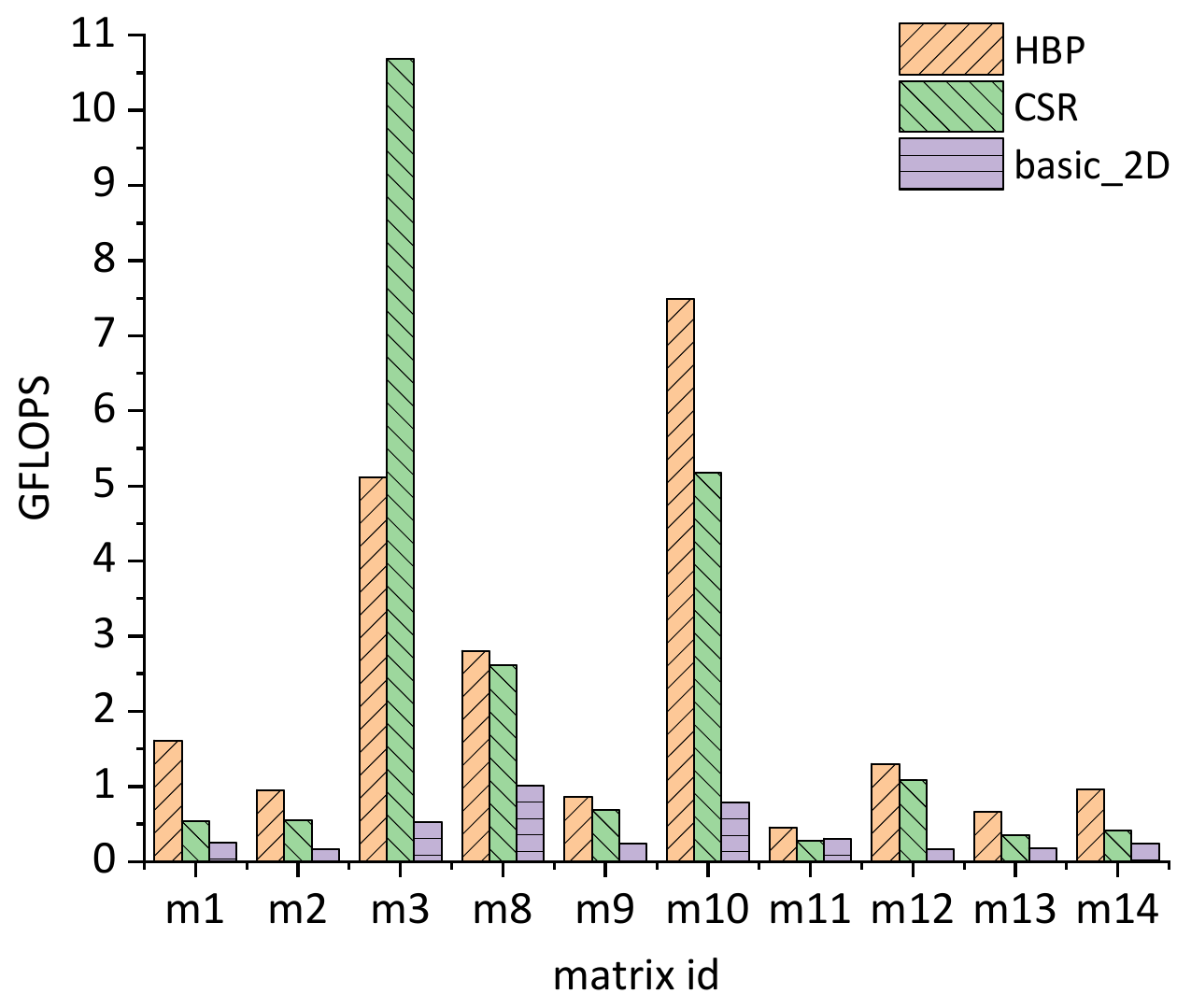}}
\caption{Evaluation of SpMV on Nvidia RTX 4090.}
\label{fig10}
\end{figure}

The process of converting the original storage format of the matrix to the HBP format we designed requires several times the original storage. Therefore, a single RTX 4090 cannot handle matrices from $m4$ to $m7$. As a result, we use other matrices to continue our testing. As shown in Fig. 10, our method achieves a maximum speedup of 3.01x (with an average speedup of 1.61x) compared to the CSR method and a maximum speedup of 9.71x (with an average speedup of 5.49x) compared to the 2D-partitioning method on Nvidia RTX 4090.

For matrices with serious load imbalance issues and scattered vector access locations, which lead to low SpMV speeds, our method achieves significant acceleration. However, the SpMV speed of the matrix $m3$ is inherently limited by the processor performance. Therefore, regardless of whether it runs on Nvidia Jetson AGX Orin or RTX 4090, the SpMV computation speed is inferior to that of the CSR format, and the high performance of RTX 4090 actually amplifies this issue.

\begin{table}[b]
\caption{THE MEM BUSY AND MEM THROUGHPUT IN RTX 4090}
\begin{center}
\begin{tabular}{|c|c|c|p{2cm}<{\centering}|p{2cm}<{\centering}|}
\hline
\multirow{2}{*}{\textbf{Matrix Id}} & \multicolumn{2}{c|}{\textbf{Mem Busy}} & \multicolumn{2}{c|}{\textbf{Mem Throughput (Gbyte/second)}}\\
\cline{2-5} & CSR & HBP & CSR & HBP \\
\hline
m1 & 0.15\% & 3.58\% & 2.85 & 145.12 \\
\hline
m2 & 0.10\% & 4.22\% & 3.29 & 189.77 \\
\hline
m3 & 5.56\% & 2.64\% & 113.3 & 123.88 \\
\hline
m8 & 0.67\% & 0.31\% & 19.05 & 15.11 \\
\hline
m9 & 0.92\% & 4.80\% & 25.53 & 215.11 \\
\hline
m10 & 14.35\% & 4.08\% & 263.69 & 169.54 \\
\hline
m11 & 0.18\% & 4.72\% & 5.26 & 211.19 \\
\hline
m12 & 0.19\% & 4.00\% & 5.2 & 178.26 \\
\hline
m13 & 0.11\% & 2.70\% & 3.15 & 121.12 \\
\hline
m14 & 0.13\% & 2.92\% & 2.67 & 128.42 \\
\hline
\end{tabular}
\label{tab2}
\end{center}
\end{table}

Using NVIDIA Nsight Compute to analyze the performance advantages of our method compared to the CSR method on Nvidia RTX 4090. The results are shown in Table 2. In the SpMV section, although prefetching the required vector segments for matrix blocks into shared memory and the HBP format itself result in transferring a considerable amount of unnecessary data, leading to increased storage operations, this contiguous read-write pattern can effectively leverage memory access optimization strategies. Consequently, the average memory throughput rate increases from 2.85 Gbyte/s to 145.12 Gbyte/s in the matrix $m1$.

\section*{Discussion and Further Work}

The hash function achieved good results in the example we provided, yet there is still considerable room for improvement in the hash function itself. This can be achieved by designing a more balanced load distribution and devising search strategies after collisions. However, the time cost of a more complex hash function itself and the atomic operations used to handle collisions are factors that need to be considered. Achieving a balance between time cost and the optimization effect of reordering remains a challenging task.

Furthermore, as mentioned previously, our method does not optimize the merging part of the 2D-partitioning approach. However, for larger-scale matrices, this inevitably limits the ultimate optimization effectiveness. We attempted to directly write the results into the result vector after the SpMV computation for each matrix block, instead of writing into an intermediate result vector. To obtain correct results, the atomicity of the writing step must be guaranteed. Unfortunately, after practical testing, we found that the cost introduced to achieve atomicity was greater than the cost of the merging step. Nevertheless, the generated intermediate vectors also exhibit strong sparsity, which suggests that threads are not fully utilized during the merging step. Therefore, optimization methods targeting this part will further enhance the speed of SpMV for large-scale matrices, and these methods can be combined with our approach to achieve more efficient optimizations.

\section*{Acknowledgment}

This work is partially supported by the Chinese Academy of Sciences Project for Young Scientists in Basic Research (YSBR-107).

\end{document}